# The Laureates of the Nobel Prize in Physics 2022
# Some personal memories

Reinhold A. Bertlmann, Faculty of Physics, University of Vienna, Boltzmanngasse 5, 1090 Vienna, Austria

The Nobel Prize in Physics 2022 was awarded to Alain Aspect (France), John F. Clauser (USA) and Anton Zeilinger (Austria) *"on the grounds of experiments with entangled photons, establishing the violation of Bell inequalities and pioneering quantum information science".* The results of their work paved the way for new technologies based on quantum information.

This is a huge triumph for a field that was once so stigmatized.

**Prologue**

It is hard for young physicists today to imagine just how frowned upon it once was to discuss questions related to the foundations of quantum mechanics. The debates in the 1930s between the great old men, Einstein and Bohr, were considered to be purely philosophical and as having no influence on actual physics. We were told that Bohr was right and Einstein was wrong. David Mermin once coined the phrase *"Shut up and calculate!"* to summarize the Copenhagen-type views of that time.

In particular, the significance of the Einstein-Podolsky-Rosen (EPR) Gedanken-Experiment of 1935 [1] was not recognized at all. It was considered to be of no use and was entirely disregarded for about 30 years. I remember that in the early 1980s Abraham Pais, a distinguished physicist from Rockefeller University who had just published the bestseller *"Subtle is the Lord: The Science and the Life of Albert Einstein"* [2], told me: *"The EPR paper was the only slip Einstein made!"* How very wrong some judgements and prophecies within the field of physics can be!

In 1964, when John Bell was on sabbatical in the US, he had the leisure to reconsider the EPR case and he wrote the paper *"On the Einstein-Podolsky-Rosen Paradox"* [3], which contained Bell's inequality. A Bell inequality, quite generally, is an inequality between expectation values of joint measurements of two parties, customarily called Alice and Bob, which all local realistic theories have to fulfil but is violated by quantum mechanics. We also speak of *Bell's Theorem*, realized via Bell inequalities: *"Local realistic theories are incompatible with quantum mechanics."* However, for a long time, Bell's work did not arouse any interest.

**John Clauser**

The first person to become interested in the subject was John Clauser, a young postdoc from Columbia University, in the late 1960s. When he studied Bell's inequality paper he saw that it contained a bound for all hidden variable theories – which he believed in – and he was fascinated and wanted to show evidence for it. So he planned to perform the experiment.

However, the value of experiments of this type was not recognized at that time. When Clauser had an appointment with Richard Feynman at Caltech to discuss an experimental EPR configuration for testing the predictions of quantum mechanics, Feynman immediately threw him out of his office, saying [4,5]:
*"Well, when you have found an error in quantum-theory's experimental predictions, come back then, and we can discuss your problem with it."*

Fortunately, Clauser remained resolute and was determined to complete the experiment. He wrote letters to David Bohm, Louis de Broglie and John Bell – all were declared realists – seeking advice or moral support. Let's quote Bell's reply [4]:
*"In view of the general success of quantum mechanics it is very hard for me to doubt the outcome of such experiments. However, I would prefer these experiments, … , to have been done … Moreover, there is always the slim chance of an unexpected result, which would shake the world!"*

Belonging to the rebellious Hippie generation of the 1960s [6], Clauser certainly *"wanted to shake the world"* [5], and in 1969 he sent an abstract to the Spring Meeting of the American Physical Society proposing the experiment. Soon afterwards, Abner Shimony called him and told him that he and his student Michael Horne had had very similar ideas. So they joined forces, and together with Richard Holt, a PhD student working with Francis Pipkin from Harvard, they wrote the famous CHSH paper [7], in which they proposed an inequality that was well adapted to experiments.

Clauser finally carried out the experiment in 1972, together with Stuart Freedman [8], a graduate student at Berkeley who received his PhD for this experiment. As pointed out in the CHSH paper [7], pairs of photons emitted in an atomic radiative cascade would be suitable for a Bell inequality test. Clauser and Freedman chose calcium atoms pumped by lasers, where the excited atoms emitted the desired photon pairs. The signals were very weak at that time, a measurement lasted for about 200 hours. For comparison with theory a very practical inequality was used, which was derived by Freedman [9]. The outcome of the experiment is well known: they obtained a clear violation of the Bell inequality very much in accordance with QM. This result was confirmed by subsequent experiments [10, 11].

Performing this experiment was truly a heroic act at that time; everything was self-made, not only the laser but also all the remaining equipment. Furthermore, Clauser could only work on the experiment because Charles H. Townes was intrigued by Clauser's ideas and offered him a job, half for Clauser's project and the other half for Townes' radio astronomy. Regrettably, because of this experiment, Clauser was not able to pursue an academic career. But it can be seen as redemption that he has now been awarded the Nobel Prize.

**Alain Aspect**

Alain Aspect, a young French physicist, was so impressed by Bell's inequality paper that he immediately decided to focus his Thèse d'Etat on this fascinating topic. He visited John Bell at CERN to discuss his proposal. John's first question to him was, as Alain later told me, "*Do you have a permanent position?*" Bell was so scared that it would ruin Aspect's career. Only after Aspect's answer in the affirmative could the discussion begin. Aspect's goal was to include variable analyzers in the setup. In the early 1980s Aspect and his collaborators performed a whole series of experiments [12, 13, 14,], with the result that the Bell inequalities used were significantly violated in each experiment. It was this that has now been recognised with the awarding of the Nobel Prize. An appreciation of Aspect's work can be found in a separate article.

**Anton Zeilinger**

In the 1990s, after Aspect's experiments, the physics community began to notice the importance and impact of such Bell-type experiments. Quantum information, communication and computation, centred around Bell inequalities and quantum entanglement, were gaining increasing interest. Now at last the prevailing attitude towards the foundations of quantum mechanics was changing. At the same time, the technical capabilities, the electronics and the lasers, were also improving considerably. Most important was the invention of a new source for creating two entangled photons, namely spontaneous parametric down conversion. Here a nonlinear crystal was pumped with a laser and the pump photon was converted into two photons that propagated on two different cones. On one cone the photons were vertically polarized and on the other horizontally. In the overlap region they were entangled. Such an EPR source was used by Anton Zeilinger and his group when they performed their celebrated experiments. But let me describe everything in order.

First of all, writing about Anton Zeilinger is a quite complex task. His accomplishments are substantial and influential in so many areas, not only in quantum physics and quantum information but also in teaching and in science administration. Zeilinger contributed to the popularisation of science – he is known by the public as "*Mister Beam*". All of this, alongside his love for philosophy and art, makes Zeilinger an incarnation of a Renaissance Scholar. So I can only focus here on just some of his achievements.

Zeilinger's interest in physics was always driven by curiosity. Even in experiments that have made applications possible, his interest has been rooted in curiosity. Also notable is the courage he showed when, in around 1990, he switched from neutron to photon physics when he became a professor at the University of Innsbruck. This was, of course, not without considerable risk. But thanks to his unerring intuition in physics and also his charisma, Zeilinger gathered highly talented students around him with whom he performed fascinating experiments.

I first met Zeilinger in person in 1991 at the Cesena Conference [15]. We immediately found common interests and decided to work together. One of our aims was to educate the new generation of young physics students on the topic of Bell-like experiments.

In 1994 we established a course titled "*Foundations of Quantum Mechanics*" at the University of Vienna. Initially it took place in a small barrack in the old AKH (Allgemeines Krankenhaus - Vienna's former general hospital, now a campus of the University of Vienna) next to the so-called Narrenturm (fools tower), a notorious place. Anton and his group would visit from Innsbruck four times per semester. The students were required to jointly deliver a talk, supplemented by a handout that would be distributed beforehand. There was always a break midway through the session, when coffee and homemade cakes would be served. It was in this cosy atmosphere that the Viennese students animatedly discussed the newest experiments being performed in Innsbruck.

Experiments which have since become famous were reported there first-hand, for example: Dik Bouwmeester, a member of Zeilinger's group, reported on "*Experimental quantum teleportation*" [16], and Gregor Weihs explained the Bell-type experiment "*Violation of Bell's Inequality under Strict Einstein Locality Conditions*" [17].

As you can imagine, the course became highly popular among the students and it continued running for about 25 years. It was undoubtedly one of the most influential courses at the Faculty of Physics in Vienna.

In 1999 Zeilinger and his whole team relocated from Innsbruck to the University of Vienna, where he became Professor of Experimental Physics. His debut was the famous experiment "*Wave-particle duality of $C_{60}$ molecules*", which demonstrated the quantum mechanical interference effects by using big molecules such as fullerenes [18]. It opened up a new field of research, led by Markus Arndt, into interfering even much bigger molecules.

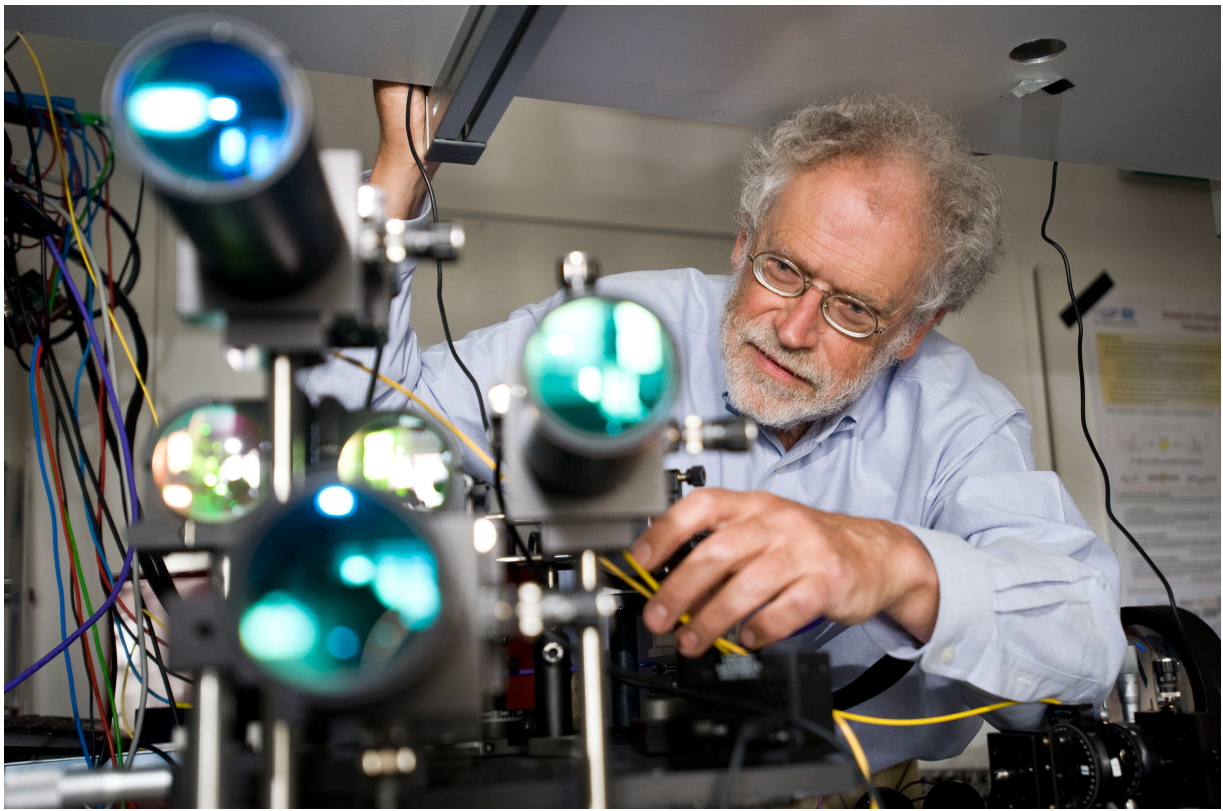

*Anton Zeilinger in his laboratory. Foto©Jacqueline Godany, IQOQI Vienna*

Next followed the "*Experimental test of quantum nonlocality in three-photon Greenberger–Horne–Zeilinger entanglement*" [19]. Back in 1989 Zeilinger had collaborated with Daniel Greenberger and Michael Horne (GHZ) and they had written one of the most influential papers in the field [20], commonly called the *GHZ Theorem*.

It was David Mermin fascinated by this issue of non-local quantum correlations, who turned the original GHZ discussion into a gedanken experiment for a system consisting of three spin-1/2 particles. In his famous Physics Today article Mermin [20] gave an extremely clear and most comprehensible presentation of the GHZ argument, which still makes it appealing to the physics community nowadays. John Bell too, who had received a copy from Mermin, replied "*I am full of admiration for your 3-spin trick*" (private communication [20]).

The *GHZ Theorem* à la Mermin asserts:
"*For a three spin-1/2 system there exists a physical situation where all local realistic theories are inconsistent and incompatible with quantum mechanics.*" It is much more restrictive than Bell's Theorem which needs expectation values, i.e., the statistics of events. These three-particle entangled states, GHZ states, opened up the entanglement research for higher multi-particle states which are important for quantum computation.

Entanglement swapping is another amazing quantum feature which Anton and his colleagues had already discovered back in 1993 [21]. It is actually the teleportation of an entangled state. More precisely, given two pairs of entangled photons, when performing a Bell state measurement for two photons of each pair, it instantaneously brings the remaining two photons into the same entangled Bell state. It was experimentally realized by Zeilinger's group in 1998 [22].

Even more amazing is the delayed-choice of entanglement swapping, where the order of the measurements is reversed as compared to standard entanglement swapping [23]. Here I became involved in Anton's activities as well and together with Walter Thirring and Heide Narhofer we were able to demonstrate that this phenomenon can be traced back to the commutativity of the projection operators of the measurements [24].

Before I continue to describe Zeilinger's experiments, I would like to mention that I also had the pleasure to collaborate with Anton on some other memorable events. In 2000, right after Zeilinger and his team had settled in Vienna, we organized a conference in honour of John Bell: "*Quantum [Un]Speakables I* " [25]. There we had numerous notable speakers, among them, just to mention a few: Mary Bell (the widow of John Bell), Gerard 't Hooft, Jack Steinberger, Roger Penrose, Alain Aspect, John Clauser, Anton Zeilinger (6 Nobel Prize winners so far), Roman Jackiw, David Mermin, Simon Kochen, Abner Shimony, Michael Horne, Daniel Greenberger, Nicolas Gisin, Helmut Rauch and others. In 2014 we organized a follow-up conference, "*Quantum [Un]Speakables II* " [26], where we had similarly distinguished speakers.

In the late 1990s there was an increasing amount of activity to test Bell inequalities. A record was set by Nicolas Gisin's group [27] in Geneva by using energy-time entangled photon pairs in optical fibers. They succeeded in separating their observers Alice and Bob by more than 10 km and could show that this distance had practically no effect on the entanglement of the photons.

In the new millennium a whole series of experiments was carried out, mainly testing the entanglement of particles at long distances via Bell inequalities. The vision was to be able, ultimately, to install a global network in outer space.

By further pushing at the limits of distance, Zeilinger's group set a record with an open-air Bell experiment over 144 km between the two Canary Islands, La Palma and Tenerife [28]. Zeilinger's former student Jian-Wei Pan and his group later extended the limit to 1120 km [29].

Up until 2015 the three loopholes: "*Locality, freedom-of-choice and fair sampling*" had only been closed separately in photon experiments. But Zeilinger's group succeeded in closing all three loopholes in one single experiment [30] (two other groups also achieved this, one in Boulder, the other one in Delft). Technically it was very challenging.

Furthermore, in 2018, an impressive "*Cosmic Bell test using random measurement settings from high-redshift quasars*" [31], whose light was emitted billions of years ago, was carried out by Zeilinger's group. This experiment pushes back to at least 7.8 Gyr ago, the most recent time by which any local-realist influences could have exploited the "freedom-of-choice" loophole to account for the observed Bell violation. Any such mechanism is practically excluded, extending from the big bang to today.

In collaboration with the Chinese Academy of Sciences, and in particular with the group of Jian-Wei Pan, Zeilinger and his team implemented quantum communication protocols between the satellite "*Micius*" and receiving stations on earth. The goal was to demonstrate a secure quantum key distribution. This key was used for the first intercontinental video call encrypted via quantum methods and demonstrated that a tap-proof quantum internet is possible.

Zeilinger also made fundamental contributions as a theorist and philosopher. Together with his student Časlav Brukner he developed a completely novel view of the meaning of a quantum state. According to their view, the quantum state represents "*information about possible future experimental outcomes*" [32,33]. Information is the most fundamental concept in quantum physics. The physical description of a system is nothing but a set of propositions together with their truth values, "*true*" or "*false*". Amazingly, relying on a few information-theoretical assumptions, they were able to derive the characteristic features of quantum mechanics such as: coherence–interference, complementarity, randomness, the von Neumann evolution equation, and, most importantly, entanglement.

I was also able to share in Zeilinger's philosophical and religious view of the world in a discussion between Anton, Walter Thirring and myself, as shown in the photo. The discussion, titled "*Zufall ist, wo Gott inkognito agiert*" (*Randomness is where God acts incognito*), was led by Thomas Kramar, a science journalist, and appeared in the Austrian newspaper "*Die Presse*" (23.3.2013).

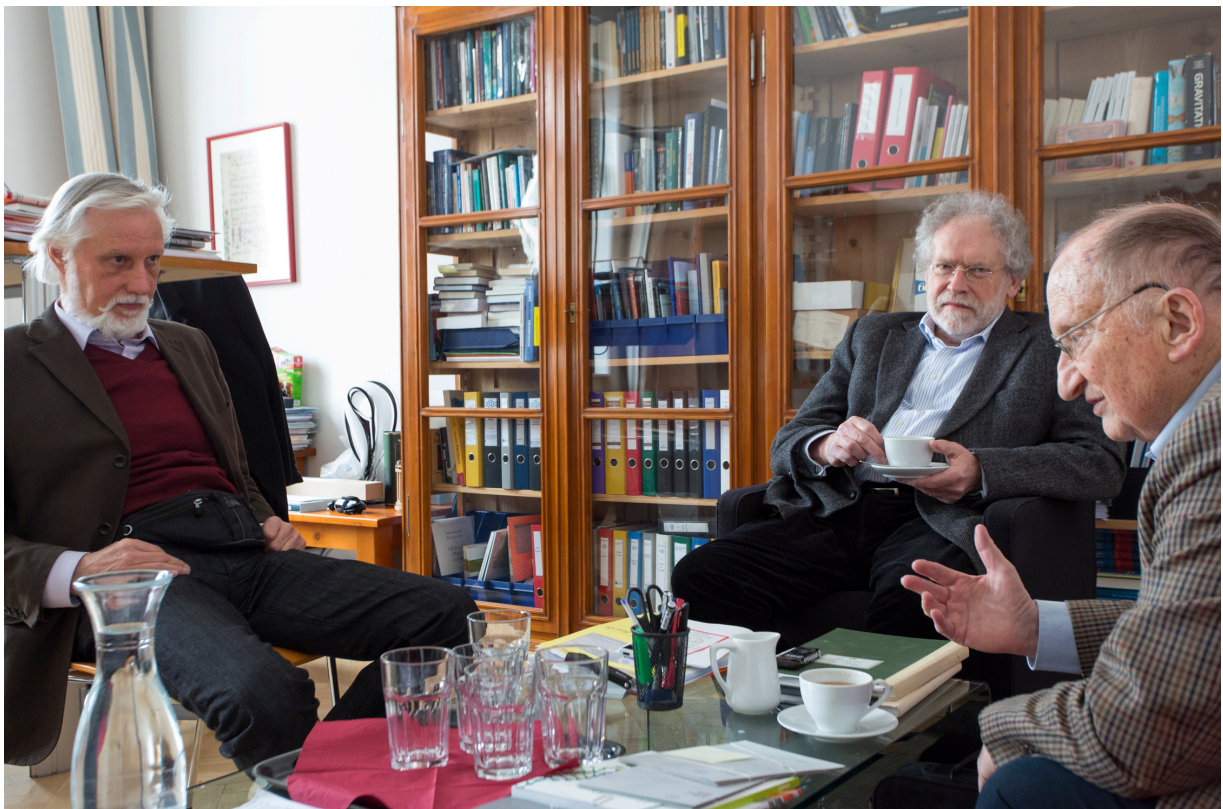

*Discussion between Reinhold Bertlmann, Anton Zeilinger and Walter Thirring on the topic "Zufall ist, wo Gott inkognito agiert" (2013). Photo ©Mirjam Reither*

Zeilinger also attracted many highly talented students who went on to become distinguished professors in their own right, reminiscent of Bohr and his group in the old days. Weihs, Arndt, Brukner and Pan I have already mentioned, some others are: Thomas Jennewein, with whom Zeilinger performed quantum cryptography experiments [34], or Phillip Walther, with whom he investigated quantum computing [35]. In the middle of the 2000s he began working with Markus Aspelmeyer on the cooling of mechanical resonators by radiation pressure [36], a new field of research now led by Aspelmeyer.

Zeilinger has held several important positions within the science administration. For instance, he became Dean of the Faculty of Physics at the University of Vienna (2006 – 2009), an important step which enabled the implementation of a reform to the law concerning the organization of Austrian universities, which he also helped to shape. Furthermore, Zeilinger was director of the Institute for Quantum Optics and Quantum Information (2004 – 2013), and he became President of the Austrian Physical Society between 1997 – 1998 and President of the Austrian Academy of Sciences between 2013 – 2022, where he was the driving force behind many innovations.

Zeilinger initiated the founding of the Institute of Science and Technology Austria (ISTA), which is an international research institute for natural and mathematical sciences located in Maria Gugging.

In 2009 Zeilinger founded the International Academy Traunkirchen, which is dedicated to supporting gifted students in science and technology. The educational platform of this Academy includes lectures that are open to the general public, as well as workshops for students and courses for school pupils with talents in the natural sciences. In the summer months the Summer Academy in Traunkirchen runs courses for artists. I also had the great pleasure of organizing several workshops with Anton for the students of our University, where they had to work on various individual topics in quantum physics. At each workshop a renowned physicist was invited to speak, for instance, in 2012 the Nobel Laureate Serge Haroche explained his "*Schrödinger cat*" cavity experiments.

Aside from the Nobel Prize, Zeilinger has received numerous international prizes and awards. Mentioning all of them would be impossible here, but some prestigious examples are: the Micius Quantum Prize (2019), the John Stewart Bell Prize (2017), the Wolf Prize (2010), and the Isaac Newton Medal (2008).

**Conclusion**

The Bell inequality experiments of the last few decades have had an enormous impact on our view of reality. In my opinion, our perception of reality must be changed radically. Objects have no properties before observation, in contrast to "*naïve*" realism, and the chronological sequence of observations is irrelevant.
Furthermore, nature is *nonlocal* as implication of Bell's Theorem. But I think, precisely this *nonlocality* feature – which deeply disturbed John Bell since for him it was equivalent to a "*breaking of Lorentz invariance*", what he hardly could accept – could be the key for a deeper and more comprehensive understanding of quantum physics.

All these experiments, which were performed for "philosophical" reasons, also triggered very practical applications, namely quantum information science, which is a prospering field today.

**The Call**

On Tuesday, October 4th, Anton Zeilinger was sitting at home, working on a paper that he and his group were due to publish. At 11 o'clock the secretary from his Institute called to tell him that there was a person on the phone who insisted on speaking to him but did not want to say why. The call came from a Swedish telephone number. When Zeilinger agreed to take the call, he found the Nobel Prize Committee on the line, who first assured him that "*It's not a fake call*" and then told him that he had been awarded the Nobel Prize in Physics along with Alain Aspect and John Clauser. Anton was speechless, overwhelmed by this wonderful recognition of his work. It came as a shock, but very much a positive one.

**Epilogue**

On a summer afternoon in 1987, John Bell and I were sitting outside in the garden of the CERN cafeteria, drinking our late 4 o'clock tea. In this relaxed atmosphere I spontaneously said to him: "*John, you deserve the Nobel Prize for your theorem.*" John, for a moment puzzled, replied with seriousness: "*No, I don't. ... it's like a null experiment, and you don't get the Nobel Prize for a null experiment. ... For me, there are Nobel rules as well, it's hard to make the case that my inequality benefits mankind.*"

But, as it turned out, in 2022 the Nobel Prize Committee did indeed conclude that the violation of Bell inequalities was of benefit to mankind and awarded the Prize to Aspect, Clauser, and Zeilinger.